

\input phyzzx
\scrollmode
\pubnum{92-43}
\titlepage

\voffset=24pt

\title{\bf PROSPECT OF HEAVY RIGHT-HANDED NEUTRINO SEARCH AT SSC/LHC
ENERGIES}
\author{Amitava Dattta ~and~ Manoranjan Guchait}
\address{Physics Department, Jadavpur University, Calcutta 700 032, India}
\andauthor{D.P. Roy}
\address{Tata Institute of Fundamental Research, Homi Bhabha Road, Bombay
400 005, India}
\abstract{Right-handed neutrinos with large Majorana mass occur naturally
in the left-right symmetric model.
We explore the prospect of such heavy Right-handed Neutrino search
Via $W_R$ decay in the Like Sign Dilepton channel at SSC/LHC.  The
standard model background can be effectively eliminated by suitable lepton
$p_T$ and isolation cuts without affecting the signal cross section
seriously.  In this way it seems possible to explore the bulk of the
parameter space $0 < M_{N_R} < M_{W_R}$, with $M_{W_R}$ going upto 3000
(2000) GeV at SSC (LHC) energy.}

\endpage

\noindent \underbar{\bf I.~Introduction}

\smallskip

The precision measurements at LEP have revealed that the Glashow--
Salam--Weinberg standard model (SM) of unified electroweak interactions is
indeed in very good agreement with the experimental data [1].  Yet there
are several aspects of the SM which are regarded as unnatural and thus
indicative of the physics beyond the SM.  For example in the SM parity
violation is introduced by hand.  Moreover the present experimental bounds
on the neutrino masses [2] indicate that these masses, even if they exist,
must be very small compared to the masses of the other fermions present in
the theory.  There is no natural explanation of such small neutrino masses
within the framework of the SM. In the left-- right symmetric (LRS) model
[3] both these problems can be taken care of naturally. In this model
parity is broken spontaneously at a high energy scale characteristic of
the masses of the $SU(2)_{R}$ gauge bosons ($W_{R}$ and $Z_{R}$). Moreover
the vacuum expectation value (VEV) of the $SU(2)_{R}$ triplet Higgs
bosons, which make the right--handed gauge bosons heavy, suppresses the
light neutrino masses [4] via the celebrated see--saw mechanism [5]. With
three generations of quarks and leptons the model contains three
heavy Majorana neutrinos of right chirality in addition to the three light
neutrinos of left chirality, which are routinely observed in low energy
leptonic and semileptonic interactions.  While the former dominantly
couples to the $W_R$, the latter couples to the standard $W_L$ boson.

An interesting experimental signature of these heavy Majorana neutrinos
($N_{\ell}, \ell=e,\mu,\tau$) is their lepton number violating
interactions.  For example, each of these neutrinos can decay into the
corresponding charged leptons of both signs with equal branching ratios
(neglecting CP violation). A characteristic of these decays is the
production of like sign dileptons (LSD)  at hadron colliders via
reactions such as
$$\eqalignno{& P P \rightarrow W_{R} \rightarrow l^{+}
N_{l} \rightarrow l^{+} l^{+} q \overline{q^{\prime}} & (1) \cr &
P P \rightarrow Z_{R} \rightarrow N_{l} N_{l} \rightarrow l^{+} l^{+} q
\overline{q^{\prime}} q^{\prime\prime} \overline{q^{\prime\prime\prime}} &
(2) \cr}
$$
The possibility of testing  one of the most sringent conservation laws of
particle physics, therefore, opens up with the advent of high energy
hadron colliders.  This was pointed out a long time back [6].
Subsequently this has been studied by several authors [7,8].  The major SM
background to the LSD signature arises  via the following reactions:
$$
PP \rightarrow Q \overline{Q} \rightarrow (q^{\prime} l^{+}
\nu)  (\overline{q^{\prime}} q_{i} \overline{q_{j}});~~~
\overline{q^{\prime}} \rightarrow l^{+} \nu  \overline{q^{\prime\prime}}
\eqno (3)
$$
where Q stands for a heavy quark (t or b or c) while the q$'$ refers to
the corresponding decay quark (b,~c~or~s) which arise due to the decays of
the heavy quarks via charged current interactions. If the produced b quark
fragments into a neutral B meson ( $B_{d}^{0}$ or $B_{s}^{0}$) then LSD's
may also arise due to $B^{0}$--$\bar B^{0}$ mixing. Obviously the
viability of detecting lepton number violation via the LSD depends on the
relative size of the signal and the background.

The cross section for the signal strongly depends on the mass of the
$W_{R}$~($M_{W_{R}}$) and the mass of the heavy neutrino (denoted
generically as $M_{N_R}$). There are several lower limits on ${W_{R}}$
derived on the basis of different assumptions: i) A strong low energy
constraint on ${W_{R}}$ comes  from the $K_L$--$K_S$ mass difference.
{}From the box diagram with both  $W_L$ and $W_R$ exchanges a lower bound of
1.6 TeV is obtained [9]. For this one has to assume the so-called manifest
LR symmetry, i.e. the Cabibbo-Kobayashi-Maskawa (CKM) mixing matrices of
the left- and right-handed sectors ($V_L$ and $V_R$) are same, which  is
rather artificial.  If  manifest LR symmetry is not assumed [10] the bound
on ${W_R}$ is relaxed; even with restrictions on fine-tunings, it can be
as low as  300 GeV [11].  ii) A  bound on $M_{W_R}$ comes from the direct
search at CDF [12]. Assuming that the new charged gauge bosons decay into
leptons and stable neutrinos of negligible masses, events with high $p_T$
electrons/muons and large missing energy are looked for. Their  lower
bound $M_{W_R} > 520$ GeV, will, however, not apply in models which we are
analysing where  the $W_R$ couples only  to heavy neutrinos  which can
decay within the detector.  iii) A bound independent of the above
assumptions come from the upper bound on $\Delta \rho$(the deviation of
the $\rho$ parameter from unity) obtained from LEP data [13].  Using the
popular choices of the Higgs fields in the literature it can be shown that
a conservative lower limit on ${W_{R}}$ is approximately 500 GeV.  Models
with  exotic  Higgs multiplets can, however,  evade this bound.  In view
of the above discussions we find it reasonable to restrict our analysis to
${W_{R}} \geq$ 500 GeV. The bound on $M_{N_R}$ is even more uncertain.
The limits from neutrinoless double beta-decay can only restrict
$M_{W_{R}}$, $M_{N_{R}}$ and $(V_{R})_{ud}$ ( the u--d element of the
right--handed quark mixing matrix)  simultaneously [11,14]. Thus the mass
limit always depends on the choice of $(V_{R})_{ud}$. However, purely from
the theoretical argument of vacuum stability one can show that
$M_{W_{R}} \geq M_{N_{R}}$ [14]. We shall restrict our analysis to this
case.

As we shall discuss below the main background after imposing suitable
kinematical cuts comes from t $\overline{t}$  production . The main
uncertainty in estimating the background, therefore, arises due to the yet
unknown top mass.  The present limits can be summarised as follows: i)
$m_{t}\geq 91$ GeV from direct mass limits from the TEVATRON [15].  ii)
The analysis of precision electroweak data from LEP yields $m_{t} =140 \pm
30$ GeV [1,16].  The observation of $B^{0}$--$\bar B^{0}$ mixing [17]
favours a large $m_{t}$ although the precise value of the lower bound is
somewhat model dependent.

The prospect of massive neutrino search via the LSD signal was considered
in ref. [8] along with the SM background.  However, their background
analysis was done for a top quark mass of 50 GeV which now appears to be
unrealistic in view of the above limits. More importantly, they could
explore only a limited range of $W_R$ and $N_R$ masses since they did not
exploit the most effective kinematical cut for the suppression of this
background, namely the one coming for lepton isolation [18]. It is
well--known that the angle between a lepton and its accompanying jet is
necessarily small for leptons arising from the decays of b and c quarks.
Since the SM background inevitably involves one such decay it is natural
to expect that it will be severely suppressed by the lepton isolation cut
while the signal remains essentially unaffected. In fact as we shall
discuss below, it is a combination of the isolation cut and the $p_{T}$
cut on the lepton which is most effective in enhancing the signal to
background ratio [19]. As a result one can probe a much larger region of
$W_{R}$ and $N_{R}$ masses than shown in [8].  This work is devoted to a
systematic exploration of these mass limits which can be probed at SSC and
LHC.

The paper is organised as follows. In sec 2 we briefly describe our
computation of the signal and background processes for the LSD cross
section.  The results are presented in sec 3. Our conclusions are
summarised in sec 4.

\bigskip

\noindent \underbar {\bf II.~ The Signal and Background Processes for LSD}

\smallskip

It has already been shown in ref 8 that the largest cross section for the
production of right-handed neutrinos and the resulting LSDs arise from the
process given in eq.  (1) of the last section.  Reactions given in eq (2)
also yield LSDs  accompanied by four jets.  However the contribution of
this process is significantly smaller  over most of the parameter space of
interest as it involves two heavy neutrinos in the final state. We shall
therefore focus our attention on reaction (1) in obtaining the region in
$M_{W_{R}}, M_{N_{R}}$ plane that can yield an observable LSD signature at
SSC or LHC.  In any case  a simultaneous analysis of both the above
reactions can only improve the conservative search limits estimated by us.
We have done a parton level Monte-Carlo calculation for the production and
decay sequence
$$
u\bar d \rightarrow W^+_R \rightarrow \ell^+ + N_\ell,~N_\ell
\rightarrow \ell^+ q\bar q' \ldots \eqno (4)
$$
where the decay processes is via virtual $W_R$ with BR$(N_\ell \rightarrow
\ell^+ q \bar{q^{\prime}}) = $ 0.5.  The relevant formulae may be found
e.g. in ref. [20]. Since the production cross section of $W_R^{+}(W_R^{-})$
involves the up (down) quark and down (up) anti-quark densities in the
proton, the former dominates over the latter by approximately a factor of
two. We have therefore computed only the former cross section.
In computing the cross section we have used the quark density functions
given in ref [21], which were parameterised using the input densities of
DFLM [22].

In order to make the search scenario not too complicated we have assumed
that each of the heavy neutrino couples dominantly with one lepton species
and have focused on like sign lepton pairs of a single species ($e^+e^+$
or $\mu^+\mu^+$).  If one assumes that $N_{\mu}$ is approximately
degenerate in mass with $N_{e}$ and considers leptons of both types then
the signal size will simply be enhanced by a factor of two.  Of course
this would not affect the signal to background ratio.

The most significant background comes from the $t\bar t$ production and
decay processes
$$
gg \rightarrow t\bar t,~~~t \rightarrow b\ell^+\nu_\ell,~~~\bar t
\rightarrow \bar b \rightarrow \bar c\ell^+\nu_\ell. \eqno (5)
$$
We have calculated this production and decay sequence using the gluon
density function of GHR [23] as input.  The resulting LSD cross-section
(after applying the kinetic cuts) shows a modest increase with the top
quark mass.  The results presented in the following section correspond to
a top quark mass $m_t = 150$ GeV.

There is a LSD background from the $b\bar b$ production, arising from the
corresponding sequential decay $b \rightarrow c \rightarrow s\ell^+\nu_e$.
However, it is practically impossible for this sequential decay lepton to
survive the large $p_T$ and isolation cuts.  A more serious background in
this case arises from $B^0$--$\bar B^0$ mixing, followed by direct
semileptonic decay of both the $\bar B^0$ particles.  We have calculated
this background assuming $B^0$--$\bar B^0$ mixing parameter $\chi \simeq
0.15$, as measured at LEP and Tevatron energies [24].  The $b\bar b$
production cross section is calculated for the next to leading order QCD
process [25]
$$
gg \rightarrow gb\bar b, \eqno (6)
$$
which dominates over the lowest order $b\bar b$ cross section in the large
$p_T$ region of our interest.  Besides, the lowest order $b\bar b$
background can be easily distinguished from the signal by the back-to-back
configuration of the decay leptons.

We have also checked the dependence of the above signal and background
cross sections on the choice of quark and gluon density functions.  Taking
these density functions from the EHLQ (set 2) parameterisation [26] gives
a somewhat larger signal cross-section while the background remains
practically the same.

\bigskip

\noindent \underbar{\bf III.~ Results}

Figs. 1--3 show the LSD signal of eq. (4) for various $W_R$ and $N_R$
masses along with the dominant background from $t\bar t$ (eq. (5)).  An
isolation cut of
$$
E^T_{AC} < 10~{\rm GeV} \eqno (7)
$$
has been applied to both the leptons, where $E^T_{AC}$ represents the
total transverse energy accompanying the lepton track within an angle of
0.4 radian [27].  Fig. 1 shows the signal and the background LSD
cross-sections against the $p^T$ of the softer lepton, $p^T_2$, for
$p^T_{1,2} > 20$ GeV.  The background decreases rapidly with increasing
$p^T_2$ for two reasons.  (1) The lepton coming from the sequential decay
$\bar t \rightarrow \bar b \rightarrow c\ell^+\nu_\ell$ has a relatively
soft $p^T$ distribution.  (2) More importantly, the isolation cut becomes
more effective with increasing $p^T$ of the lepton coming from $b$ decay,
as pointed out in [19].  Indeed, to a first approximation, there is a
kinematic bound [19]
$$
E^T_{AC} > {p^T_\ell m^2_c \over m^2_b - m^2_c} - {m^2_b - m^2_c \over
4~p^T_\ell}, \eqno (8)
$$
which implies an upper bound of $p^T_\ell < 100$ GeV for $m_b = 5~{\rm
GeV}, m_c = 1.5~{\rm GeV}$ and $E^T_{AC} < 10$ GeV.  This is reflected in
Fig. 1, where the background cross-section goes down by one order of
magnitude by increasing the lepton $p^T$ cut from 20 to 40 GeV, and by two
orders by increasing it to 60 GeV.  It may not be possible in practice to
decrease the background by two orders of magnitude due to effects like jet
fragmentation, which have not been taken into account above.  A detailed
investigation of these effects is in progress.  However, it is reasonable
to expect on the basis of existing simulations [27], that the background
can be decreased by at least one order of magnitude by increasing the
lepton  $p^T$ cut to 40 GeV, say.  As we shall see below, this will be
adequate for our purpose, since a background of this level can be
effectively eliminated by other kinematic cuts.  It should be mentioned
here that the $p^T_2 > 40$ GeV cut has little effect on the signal cross
section except when $M_{N_R} \simeq M_{W_R}$ or when $M_{N_R}$ is very
small.  In the former case the lepton produced in association with $N_R$
is too soft to survive the $p^T_2 > 40$ GeV cut, while in the latter case
the lepton coming from the $N_R$ decay does not survive the isolation cut.

Fig. 2 shows the signal and background LSD cross-sections against the
$p^T$ of the harder lepton, $p^T_1$, with the $p^T_{1,2} > 40$ GeV cut.
Although the magnitude of the background is still large compared to the
signal for most of the parameter space, the two can be easily separated
from their $p^T_1$ distributions.  While the background decreases rapidly
with $p^T_1$, the signal shows a Jacobian peak at [28]
$$
p^T_1 = \left(M^2_{W_R} - M^2_{N_R}\right) \big/ 2~M_{W_R}. \eqno (9)
$$
Thus a $p^T_1 > 200$ GeV (150 GeV) cut in Fig. 2a(b) will effectively
eliminate the background without affecting the signal.  Alternatively one
can also separate the signal and the background from the dilepton
invariant mass distribution shown in Fig. 3.  Of course we do not see any
particular reason to prefer the dilepton invariant mass distribution over
the $p^T$ distribution of the harder lepton.

We have checked that the LSD background from $b\bar b$ production and
mixing is comparable to that from $t\bar t$ for the $p^T_{1,2} > 20$ GeV
cut.  Increasing the cut to $p^T_{1,2} > 40$ GeV, however, suppresses the
$b\bar b$ background more strongly than the $t\bar t$ for the following
reason.  While for the $t\bar t$ case one of the decay leptons is hard and
isolated and hence not affected by the increasing $p^T$ cut, it affects
both the decay leptons for the $b\bar b$ case.  As a result the LSD
background from $b\bar b$ is an order of magnitude smaller than that from
$t\bar t$, for the $p^T_{1,2} > 40$ GeV cut.  Moreover the $p^T_1$
distribution in this case is even softer than that of the $t\bar t$
background shown in Fig. 2.  Therefore we have not displayed the $b\bar b$
background.

It is clear from the above discussions that the right-handed neutrino
signal can be effectively separated from the standard model background in
the isolated LSD channel, with $p^T_{1,2} > 40$ GeV.  Thus the prospect of
heavy right-handed neutrino search is essentially controlled by the signal
size in this channel.  This is shown in Table I for $M_{W_R} = 2000,~3000$
GeV (1000, 2000 GeV) at SSC (LHC) energy.  For each $W_R$ mass the signal
cross-section is shown for three representative $N_R$ masses in the range
$O < M_{N_R} < M_{W_R}$.  It is clear from this Table that one can explore
the bulk of the mass range $M_{N_R} < M_{W_R}$ with $M_{W_R}$ going upto
3000 GeV (2000 GeV) at SSC (LHC), with the expected luminosity of 10
events/fb.   With the high luminosity option of 100 events/fb, the LHC
search can also go upto $M_{WR} = 3000$ GeV.  Finally the search can be
extended to somewhat larger values of $M_{W_R}$, but only for a limited
range of $M_{N_R}$ around $M_{W_R}/2$.

The distinctive features of the signal are (1) clustering of the total
invariant mass of the $2$ jets and the $2$ leptons at $M_{W_R}$ and (2)
clustering of the invariant mass of the 2 jets with one of the leptons at
$M_{N_R}$.  The right-handed $W$ and neutrino masses can be easily
obtained from these mass peaks.  For most of the parameter space of
$M_{N_R}$ the mass peak is expected to show up in the invariant mass of
the $2$ jets with the softer lepton.  For $M_{N_R} \simeq M_{W_R}$,
however, the lepton produced in association with $N_R$ becomes softer than
the lepton from $N_R$ decay (eq. 4), so that the $N_R$ mass peak shows up
in the invariant mass of the 2 jets with the harder lepton.  This is
illustrated in Fig. 4.

\bigskip

\noindent \underbar {\bf IV.~ Summary}

In the left-right symmetric models one expects heavy right-handed
neutrinos with mass $M_{N_R} \leq M_{WR}$.  We explore the prospect of
searching for such neutrinos at SSC/LHC.  The most prominent source of
$N_R$ production is via $W_R$ decay, resulting in a characteristic
signature of like sign dileptons.  The standard model background to this
channel can be eliminated by a combination of lepton $p^T$ and isolation
cuts without any serious reduction in the signal.  Thus the search limit
is essentially controlled by the size of the signal.  One expects a viable
signal for most of the mass range $M_{N_R} < M_{WR}$ with $M_{WR}$ going
upto 3000 GeV at SSC and 2000 GeV at LHC.

\bigskip

\noindent  {\bf Acknowledgments:}

A.D. and M.G. acknowledge partial support by the Department of Science and
Technology, India.  D.P.R. acknowledges discussions with S. Uma Sankar and
Biswarup Mukhopadhyaya.

\endpage

 \centerline {\bf References}

\medskip

\item {[1]} See, e.g., J. Carter, in Proc. Intl.
Lepton--Photon Symp. and Europhysics Conference
on High Energy Physics, Geneva, 1991, Vol. 2,
(World Scientific, 1992) and references therein.

\item {[2]} Particle Data Group,Phys.Rev.{\bf 45D},S1(1992)

\item {[3]} J.C. Pati and A. Salam, Phys. Rev. {\bf D10}, 275 (1974);
R.N. Mohapatra and J.C. Pati, Phys. Rev. {\bf D11}, 366, 2588 (1975);
G. Senjanovi\'{c} and R.N. Mohapatra, Phys. Rev. {\bf D12}, 1502 (1975).

\item {[4]} R.N. Mohapatra and G. Senjanovi\'{c}, Phys. Rev.
Lett. {\bf 44}, 912 (1980); R.E. Marshak and R.N. Mohapatra, {\it ibid.}
{\bf 44}, 1316 (1980).

\item {[5]} M. Gell-Mann, P.Ramond and R.Slansky, in
`Supergravity' (eds. P.Van Nieuwenhuizen and D.Friedman, North
Holland,1979); T.Yanagida, Proc.Workshop on Unified Theory and Baryon
Number of the Universe(eds. O.Swada and A.Sugamoto, KEK, Japan,1979).

\item {[6]} W.-.Keung and G. Senjanovi\'{c}, Phys.
Rev.Lett.{\bf 50},1427(1983).

\item {[7]} E. Eichtern, I.Hinchliffe, K.Lane and C.Quigg, Rev.
Mod.Phys. {\bf 56}, 579 (1984); J.A.Griffols, A.Mendez and R.M.Barnett
Phys.Rev. {\bf D 40}, 3613(1989); F.Fergulio {\it et al}, Phys.Lett.
{\bf B 233}, 512(1989); D.A. Dicus and P. Roy, Phys. Rev. {\bf D 44}, 1593
(1991).

\item {[8]} H.Tso--hsiu, C.Cheng--rui and T.Zhi-jian,Phys.Rev.
{\bf 42}, 2265(1990).

\item {[9]]} G. Beall, M. Bander and A. Soni, Phys. Rev. Lett.
{\bf 48}, 848 (1982); G. Ecker and W. Grimus, Nucl. Phys. {\bf
B258}, 328 (1985).

\item {[10]} A. Datta and A. Raychaudhuri, Phys. Lett. {\bf B122}, 392
(1982); P. Basak, A. Datta and A. Raychaudhuri, Z. Phys. {\bf C20},
305 (1983);  F. Olness and M.E. Ebel, Phys. Rev. {\bf D30}, 1034 (1983).

\item {[11]} P. Langacker and S. Uma Sankar, Phys. Rev.  {\bf D40}, 1569
(1989).

\item {[12]} See, e.g.,  J. Freeman, in `Particle Phenomenology in the
90's', eds. A. Datta, P. Ghose and A. Raychaudhuri (World Scientific,
1992).

\item {[13]} G.Bhattacharyya, A.Datta, A.Raychaudhuri and
U.Sarkar. Pre-Print ICTP (1992).

\item {[14]} R.N.Mohapatra, Phys.Rev. {\bf D34}, 909{1986}.

\item {[15]} F.Abe {\it et al} (CDF collaboration),
Phys. Rev. Lett. {\bf 68}, 447(1992).

\item {[16]} See, e.g., J. Ellis, in proc. Intl. Lepton-Photon Symp. and
Europhysics Conf. on High Energy Physics, Geneva, 1991, Vol. 2 (World
Scientific, 1992).

\item {[17]} P.J.Franzini,Phys.Rep. {\bf 173}, 1 (1989).

\item {[18]} R.M. Godbole, S. Pakvasa and D.P.Roy, Phys. Rev. Lett.
{\bf 50}, 1539 (1983).

\item {[19]} D.P. Roy, Phys. Lett. {\bf B196}, 395 (1987).

\item {[20]} V. Barger and R.J.N. Phillips, Collider Physics,
Addison-Wesley (1987).

\item {[21]
} M.G. Gluck, R.M. Godbole and E. Reya, Dortmund Preprint,
DO-TH-89/16.

\item {[22]} M. Diemoz, F. Ferroni, E. Longo and G. Martinelli, Z. Phys.
{\bf C39}, 21 (1988).

\item {[23]} M. Gluck, F. Hoffmann and E. Reya, Z. Phys. {\bf C13}, 119
(1982).

\item {[24]} See, e.g. P. Roudeau and M.V. Danilov in Proc. Int.
Lepton-Photon Symp. and Europhysics Conf. on High Energy Physics, Geneva,
1991, Vol. 2 (World Scientific, 1992).

\item {[25]} R.K. Ellis and J.C. Sexton, Nucl. Phys. {\bf B282}, 642
(1987).

\item {[26]} E. Eichten, I. Hinchliffe, K. Lane and C. Quigg, ref 7.

\item {[27]} F. Cavanna, D. Denegri and T. Rodrigo, in Proc. LHC Workshop,
Vol. II, 329, CERN 90-10 (1990).

\item {[28]} The Jacobian peak is suppressed for $M_{N_R} \ll M_{W_R}$
(short dashed lines of Fig. 2) due to the isolation cut on the
softer lepton coming from the $N_R$ decay. Large $p^T_1$ corresponds to
large $p_T$ of the associated $N_R$ and hence a significant isolation cut
on its decay lepton.

\endpage

\hrule width 0pt

\endpage

\centerline{\bf Figure Captions}

\medskip

{\itemlength{\rm Fig.~2.~:}

\item {\rm Fig.~1.~:} The isolated dilepton ($e^+e^+$ or $\mu^+\mu^+$)
cross-section is shown against the $p^T$ of the softer lepton at (a) SSC
and (b) LHC energies.  The signal cross-sections are shown for different
choices of $W_R$ and $N_R$ mass along with the dominant background from
$t\bar t$ cascade decay.  The upper (lower) set of signal curves
correspond to the lower (higher) $W_R$ mass.

\item {\rm Fig.~2.~:} The isolated dilepton cross-section, with a $p^T >
40$ GeV cut for both the leptons, is shown against the $p^T$ of the harder
lepton at (a) SSC and (b) LHC energies.  The cross-section curves are as
in Fig. 2.

\item {\rm Fig.~3.~:} The isolated dilepton cross-section, with a $p^T >
40$ GeV cut on both the leptons, is shown as a function of the dilepton
invariant mass at (a) SSC and (b) LHC energies.  The cross-section curves
are as in Fig. 2.

\item {\rm Fig.~4.~:} The distribution of the signal cross-section in the
invariant mass of the two jets and the softer lepton (harder lepton for
$M_{N_R} \simeq M_{WR}$, shown by the long dashed curves) at LHC energy.

}

\endpage

\bye